\documentstyle[sprocl,psfig]{article}

\def\Journal#1#2#3#4{{#1} {\bf #2}, #3 (#4)}

\def\be{\begin{equation}}
\def\ee{\end{equation}}
\def\bea{\begin{eqnarray}}
\def\eea{\end{eqnarray}}

\begin{document}

\title{VLT/UVES OBSERVATIONS OF THE DLA AT Z=3.39\\ TOWARDS 
QSO 0000-2621}

\author{P. Molaro, P. Bonifacio, M. Centuri\'on,  G. Vladilo}

\address{Osservatorio Astronomico di Trieste, Via G.B. Tiepolo 11, 
\\ I 34131, Trieste Italy\\} 

\author{S. D'Odorico}

\address{European Southern Observatory, Karl-Schwarzschild-Str. 2\\
D-85740 Garching Germany\\}

\author{ S. A. Levshakov,}

\address{Department of Theoretical Astrophysics, Ioffe 
Physico-Technical Institute\\
194021 St Petersburg, Russia\\}

\maketitle\abstracts{ 
UVES observations at  ESO/VLT of the DLA at z=3.39 towards QSO
0000-2621 provided refined  abundances for  Fe, Si, N, Ni,
and new abundances for  O, Ar, P, Zn, Cr, H$_{2}$.
These determinations confirm that the system has remarkable properties with
 the lowest metallicity, a very low H$_{2}$ molecular fraction
and  it has no   detectable presence of dust. 
The  
$[\alpha/Fe-peak]$ ratio is $\approx 0.1 $,  
 challenging current models of chemical evolution.
The O/Ar and O/S ratios are the
 same as  those observed in BCGs, suggesting a universal IMF. }

\section{Chemical abundances}

UVES  observations of  QSO 0000-2621 (V=17.5 mag, z$_{em}$=4.108), which  
shows a 
DLA system at  z$_{abs}$=3.3901,
  have been acquired 
 at  ESO/VLT Kueyen 8.2m during  commissioning
 \cite{m2000}. 
 
\begin{figure}
\centerline{\psfig{figure=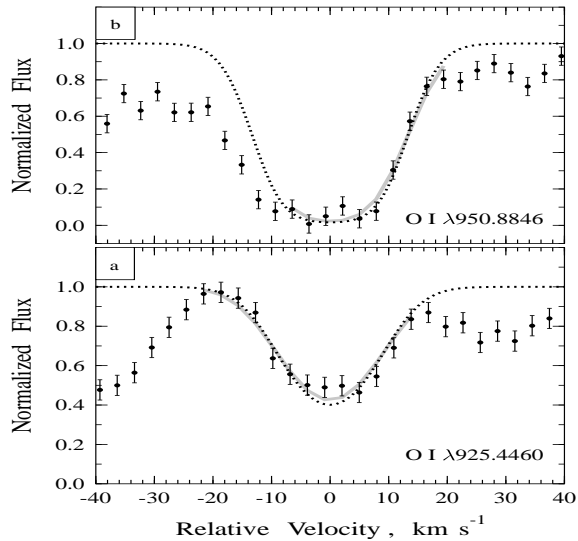,height=7.1cm,width=7.5cm,clip=t}}
\caption{ Oxygen lines}
\end{figure}
 
The Fe abundance is  found  [Fe/H]=-2.04,   Cr is  
[Cr/H]=-2.01 while  Ni is    [Ni/H]=-2.27, 
somewhat below the other Fe-elements.
The undepleted  Zn is measured at  [Zn/H]=-2.07 and so far it is the 
lowest in all  DLAs showing that this  material  has indeed been processed
very little. The 
 Zn  measure is also the one achieved  at the highest redshift 
 and suggests
a  mild chemical evolution when compared with extant data for Zn in the DLAs.
However,   no   evolution is found when 
 the column-density-weighted mean is considered instead \cite{v2000}.
 Depleted elements such as Fe and Cr are at the same level of
 the undepleted Zn, showing that there  is no detectable presence of
dust.
Si is found at 
[Si/H]=-1.91 or
[Si/Fe]=0.13 $\pm$ 0.04,  
at variance with the  mean value    [Si/Fe]=+0.43$\pm$ 0.18 
found in  DLAs.

A special effort has been made to measure abundances by using transitions
which fall within  the Ly$\alpha$  forest where  
HI contamination and   uncertainties in the  continuum location make it
a   challenging task.
The 
  lines within the Ly$\alpha$ forest are  fitted simultaneously  
 with  those  outside the forest leaving the local continuum  free
 to vary \cite{m2001}. With this procedure we 
 have been able
 to  measure the abundances of O, Ar,  P and H$_{2}$.

In the DLAs
the  OI $\lambda$ 1302 \AA\ is always found saturated, while the
 OI $\lambda$ 1356\AA\ is never detected. In our spectrum we succeed in
 detecting 7 OI lines mixed in the forest. The 
  best cases for the elemental measure are the 
 OI $\lambda$ 950  and  $\lambda$ 925 \AA\  shown in the figure.
   The O abundance is 
found [O/H]=-2.01. Note  that this value has been slightly revised with 
comparison to the earliest measure \cite{m2000} and 
gives better
matching between the two transitions involved.

  For the first time  we detected the 
ArI $\lambda$ 1048 and $\lambda$ 1066 \AA\ lines \cite{m2001} which yield an abundance
of [Ar/H]=-1.91. Argon is both an $\alpha$ element 
and  a very  volatile one.
 We detected the PII 963 \AA\ line \cite{m2001} obtaining 
  [P/H]=-2.31. 
$^{31}$P is an   odd-element
 likely produced by $n$-capture in C and Ne shell burning and 
 non-depleted in the ISM. 
This measure  is unique  at low metallicity and  the   
[P/Fe]=-0.2 ratio implies   metallicity-dependent  yields.

Molecular hydrogen has been detected \cite{l2001} from 
 the L(4-0)R(1) and W(2-0)R(0) lines yielding a molecular
        fraction of $f(H2) = 6.8 \pm 2.0 \times 10^{-8}$ 
         which is $\approx$  100 times lower than 
          the molecular fraction in the primordial gas at
        z $\le$ 50.
  A   correlation  has been found between $f(H_2)$ and [Cr/Zn] 
  in the DLAs, suggesting 
 a  negligible dust condensation in the z=3.39 absorber \cite{l2001}.
  
\subsection{The $\alpha$-elements/Fe-peak ratio and the IMF}

Comparison of   $\alpha$ elements with zinc, as a proxy for the 
 iron peak elements, gives  
[O/Zn] = 0.06$\pm$ 0.07
and [Ar/Zn]=0.16$\pm$ 0.06. 
 S  has also been measured \cite{l} providing
  [S/Zn]=0.16$\pm$ 0.06. 
The average of all  four $\alpha$ elements at our disposal
with the  Fe-peak elements available,  without  considering
 Ni which is found somewhat below, is
$\mathrm [O, Si, S, Ar / Zn, Fe, Cr]$ = 0.10 $\pm$0.06
and confirms  early observations \cite{m96}.
Although similar analysis are not presently available for other
DLAs,  there are indications that this result may be rather general.
[S/Zn] $\approx$ 0 in the small sample  where it has been measured     
\cite{cbmv}. [Si/Zn] $\approx$ 0 when Si  has been corrected for 
  the presence of dust 
   \cite{v98}, and  low $\alpha$ abundances
   are found in 
  low-dust systems \cite{pesb}. 

Low $\alpha$ abundances  at [Fe/H]$\approx$ -2 are not easy to explain.
Burst(s) of
star formation followed by quiescent periods 
where Type I SNe evolve have been considered \cite{mmv}.
 However, for  SN type I lifetimes  of  $\approx$ 1 Gyr to produce
effects   at z=3.4 we need   stars formed at a very 
early time, namely  z$>$7.

For the first time  in a DLA system we have at our disposal information for O, Ar
and S abundances. 
The S/O \& Ar/O ratios measured in a variety of celestial objects
are found to be independent from metallicity, which has been taken as 
evidence for
a   universal IMF \cite{hw}.
In our DLA we find 
 $\log$(Ar/O)=-2.25 $\pm$ 0.06 and 
$\log$(S/O)=-1.57$\pm$ 0.06. These elemental ratios are
 almost identical to  the -2.25 $\pm$0.09  
 and -1.55 $\pm$0.09   values   measured respectively in  
 Blue Compact Galaxies \cite{it}. 
Thus  the S/O \& Ar/O constancy  is probed for the first time
in a possibly different type
of objects and for  a much lower O abundance than in the BCGs.


\section*{References}
\begin{thebibliography}{99}
\bibitem{m2000}Molaro P., Bonifacio P., et al
\Journal{\em ApJ}{ 541}{ 54}{2000}
\bibitem{v2000}Vladilo G., Bonifacio P., Centuri\'on M., Molaro P.,
  \Journal{\em ApJ}{ 543}{ 24}{2000}
\bibitem{m2001}Molaro P., Levshakov S.A., et al
 \Journal{\em ApJ}{}{}{2001} in press astro-ph/0010434
\bibitem{l2001}Levshakov S.A., Molaro P., et al
\Journal{\em  astro-ph/0011470}{}{}{2001}
\bibitem{l}Lu L.,  Sargent W. L. W., et al
\Journal{\em ApJS}{ 107}{ 475}{ 1996}
\bibitem{m96} Molaro P., D'Odorico S., et al
\Journal{\em A\&A}{ 308}{ 1}{1996}
\bibitem{cbmv}Centuri\'on M., Bonifacio P., Molaro P., Vladilo G., 
\Journal{\em ApJ}{ 536}{ 540}{1999} 
\bibitem{v98} Vladilo G.,  \Journal{\em ApJ}{  493} {583}{1998}
\bibitem{pesb} Pettini M., Ellison S., L., et al 
\Journal{\em ApJ}{ 510}{ 576}{1999}
\bibitem{hw}Henry R. B. C., Worhtey G., 
\Journal{\em PASP}{ 111}{ 919}{1999}
\bibitem{it}Izotov Y., \& Thuan T., X.,
\Journal{\em ApJ}{ 511}{ 639}{1999}
\bibitem{mmv}Matteucci F., Molaro P., Vladilo G., 
 \Journal{\em A\&A} {321}{ 45}{1997}






\end{thebibliography}
\end{document}